\documentclass[prb,aps,twocolumn,showpacs,preprintnumbers,superscriptaddress,amsmath,amssymb]{revtex4-1}
\usepackage{graphicx}% Include figure files
\usepackage{dcolumn}% Align table columns on decimal point
\usepackage{bm}% bold math
\usepackage{color}

\begin{document}

\title{Orbital-dependent correlations in PuCoGa$_5$}

\author{W. H. Brito}
\address{Condensed Matter Physics and Materials Science Department, Brookhaven National Laboratory, Upton, New York 11973, USA.}
\author{S. Choi}
\address{Condensed Matter Physics and Materials Science Department, Brookhaven National Laboratory, Upton, New York 11973, USA.}
\author{Y. X. Yao}
\address{Ames Laboratory-U.S. DOE and Department of Physics and Astronomy, Iowa State University, Ames, Iowa 50011, USA.}
\author{G. Kotliar}
\address{Department of Physics and Astronomy, Rutgers University, Piscataway, New Jersey 08854, USA.}
\address{Condensed Matter Physics and Materials Science Department, Brookhaven National Laboratory, Upton, New York 11973, USA.}

\begin{abstract}

We investigate the normal state of the superconducting compound PuCoGa$_5$ using the combination of density functional theory (DFT) and dynamical mean field theory (DMFT), with the continuous time quantum Monte Carlo (CTQMC) and the vertex-corrected one-crossing approximation (OCA) as the impurity solvers. 
Our DFT+DMFT(CTQMC) calculations suggest a strong tendency of Pu-5$f$ orbitals to differentiate at low temperatures. The renormalized 5$f_{5/2}$ states exhibit a Fermi-liquid behavior whereas one electron in the 5$f_{7/2}$ states is at the edge of a Mott localization. We find that the orbital differentiation is manifested as the removing of 5$f_{7/2}$ spectral weight from the Fermi level relative to DFT. We corroborate these conclusions with DFT+DMFT(OCA) calculations which demonstrate that 5$f_{5/2}$ electrons have a much larger Kondo scale than the 5$f_{7/2}$.

\end{abstract}

\maketitle

\section{Introduction}

Orbital-dependent correlations have emerged as a key concept to understand the physics of a large number of materials.  
Early on Anisimov and coworkers~\cite{anisimov_ruth} suggested an orbital selective Mott transition in the ruthenates.
Later, orbital-dependent correlations were observed in the normal state of iron-based superconductors.~\cite{ziping,miao} More recently, 
orbital differentiation has also been shown to play an important role in the 5$f$ manifold of UO$_2$.~\cite{werner_uo2,lanata_uo2}
In this paper, we point out that orbital differentiation also occurs in Pu-5$f$ systems, by presenting a study of PuCoGa$_5$. This suggests that orbital differentiation is a very general phenomena in multiorbital systems.

Among the group of Pu-based compounds, PuCoGa$_5$ has attracted major interest since its superconductivity develops at $T_{c}$ = 18.5 K,~\cite{sarrao1} which is the record transition temperature among the family of heavy fermion superconductors.~\cite{sarrao_review} Moreover, its superconducting properties indicate the existence of heavy quasiparticles~\cite{javorsky} while its normal state exhibits a non-Fermi liquid resistivity up to 50 K.~\cite{wastin}
The complexity of elemental plutonium is also seen in the properties of PuCoGa$_5$. Analogous to what happens in $\delta$-Pu, neutron scattering measurements pointed out the absence of localized magnetic moments in the normal state,~\cite{hiess1} which indicates an unconventional electron pairing mechanism. In fact, a comparison between the properties of PuCoGa$_5$ and PuCoIn$_5$ has suggested two distinct electron pairing mechanisms, one due to spin fluctuations and the other mediated by valence fluctuations.~\cite{bauer1,kout}
Although the electron pairing mechanism is still under debate, more recent experiments clearly evidence a d-wave superconductivity in PuCoGa$_5$.~\cite{daghero}

Early theoretical works employed different methods to study the electronic structure of PuCoGa$_5$. Density functional theory (DFT) calculations showed that states around the Fermi level come mainly from Pu-5$f$ states and that the paramagnetic Fermi surface is essentially two-dimensional.~\cite{opahle} 
However, these calculations fail to describe the magnetic ground state, which was predicted to be antiferromagnetic or ferromagnetic due to the nearly identical total energies of both phases. This issue was later solved by LSDA+U calculations which predicted Fermi surfaces very similar to the ones obtained within DFT.~\cite{oppeneerLDAU}
The normal state of PuCoGa$_5$ was also studied using the combination of DFT with dynamical mean field theory (DMFT). 
By means of DFT+DMFT calculations using the spin-orbit T-matrix and fluctuating exchange (SPFT) approximation, Pourovskii \textit{et al.}~\cite{pourovskii} obtained a nonmagnetic state with van Hove singularities in the spectral function at 500 K. Furthermore, the authors pointed out that these singularities can result in a strong $\mathbf{q}$ dependence of the magnetic susceptibility, which advocates to a d-wave superconductivity mediated by spin fluctuations. Moreover,  DFT+DMFT calculations using the vertex-corrected one-crossing approximation (OCA) were used to compare the correlation effects in PuCoGa$_5$ to PuCoIn$_5$.~\cite{zhu} In particular, the authors found a three-peak structure in the Pu-5$f$ density of states for both materials,  wherein the central peak is associated with strongly renormalized quasiparticles.

With this motivation we reconsider the issue of orbital differentiation in PuCoGa$_5$ using the DFT+DMFT method~\cite{review2} employing state of the art impurity solvers. 
We find strong orbital differentiation in this material, with the 5$f_{7/2}$ states more renormalized than the 5$f_{5/2}$ states,  and equivalently  the coherence scale of  
the 5$f_{7/2}$ states much smaller than the one of the 5$f_{5/2}$ states. These conclusions were obtained  using both  CTQMC and OCA as impurity solvers, and  hence orbital differentiation is a  robust property of this material, which had not been discussed previously in the literature.

\section{Computational Methods}
\label{method}

Our calculations were performed using the fully charge self-consistent DFT+embedded-DMFT approach,~\cite{hauleWK} as implemented in K. Haule's code.~\cite{haulepage}
The DFT calculations were performed within Perdew-Burke-Ernzehof generalized gradient approximation (PBE-GGA),~\cite{pbe} as implemented in Wien2K package.~\cite{wien} To solve the DMFT effective impurity problem we used the Continuous time quantum Monte Carlo (CTQMC) method~\cite{ctqmc} and the vertex-corrected one-crossing approximation (OCA).~\cite{pruschke} In particular, we use the same values for the on-site Coulomb repulsion $U = 4.5 $ eV and Hund's coupling $J = 0.512$ eV which were used to describe the ground state of $\delta$-Pu.~\cite{janoschek} For the double-counting correction term we use the standard fully localized-limit form~\cite{anisimovEdc} with $n_{f}^{0}$ = 5.2, which is the average occupancy of Pu-5$f$ states in the $\delta$-Pu as reported in Ref.~\onlinecite{lanata} .

\section{Results and Discussions}

\begin{figure}[!htb]
\includegraphics[scale=0.38]{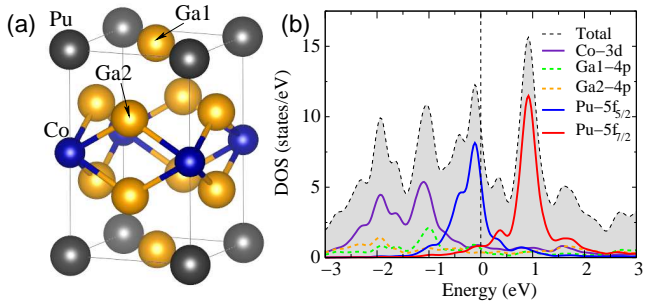}
 \caption{(a) Crystal structure of PuCoGa$_5$ (space group $P4/mmm$). Plutonium, cobalt and gallium atoms are represented by black, blue and yellow spheres, respectively. In (b) we show the DFT calculated density of states. Shaded region indicates the total density of states while the lines in blue, red, indigo, dashed green, and dashed orange denote the Pu-5$f_{5/2}$, Pu-5$f_{7/2}$, Co-3d, Ga1-4p, and Ga2-4p projected density of states, respectively. The Ga-4p projected density of states were multiplied by a factor of 5 for clarity.}
\label{fig:fig1_dosdft}
\end{figure}

Similar to Ce-115 materials, PuCoGa$_5$ crystallizes in a HoCoGa$_5$ tetragonal structure, which can be viewed as composed of PuGa$_3$ and CoGa$_2$ layers, as shown in Fig.~\ref{fig:fig1_dosdft}(a). As pointed out by Sarrao \textit{et al.},~\cite{sarrao_review} an interesting feature among the family of Pu-115 superconductors is that the T$_c$ is directly connected with the distance between these layers, with PuCoGa$_5$ being the member with highest T$_c$ and smaller lattice constant $c$.
In our calculations we used the experimental lattice structure, with $a = 4.2$ \AA{} and $c = 6.8$ \AA{} as reported in Ref.~\onlinecite{hiess}. In Fig.~\ref{fig:fig1_dosdft}(b) we show the calculated DFT(GGA) total and projected density of states.

Our DFT calculations indicate that bands near the Fermi level are mainly of Pu-5$f$ character, where the peak just below $E_{f}$ corresponds to 5$f_{5/2}$ states, while the peak around 1 eV to 5$f_{7/2}$ states, which agrees with previous DFT calculations.~\cite{zhu} The Co-3d states give rise to peaks centered at -1.1 and -1.9 eV. The contribution of Ga-4p states is rather small from -3 to 3 eV around $E_f$.

\begin{figure*}[!ht]%[!htb]
\includegraphics[scale=0.53]{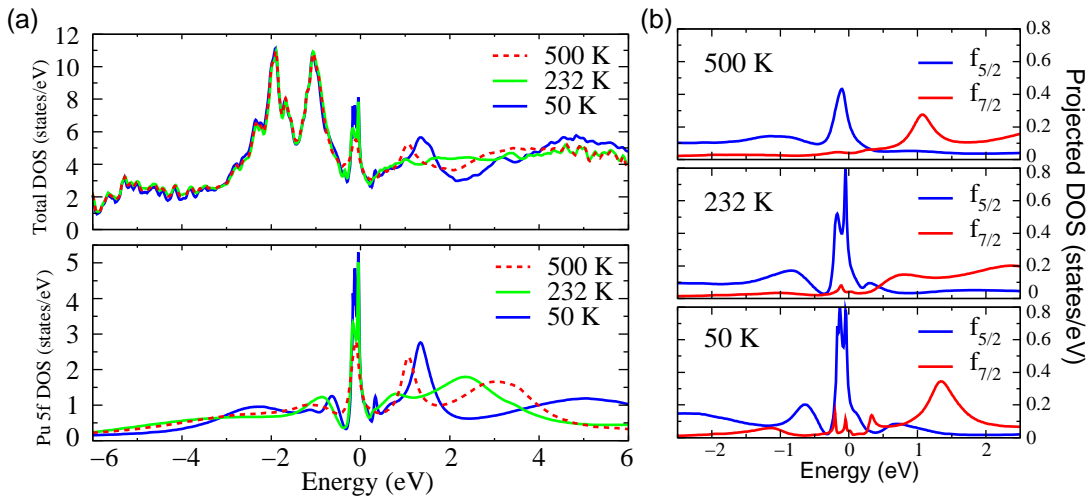}
 \caption{(a) DFT+DMFT based total (upper panel) and Pu-5$f$ (lower panel) projected density of states at 500 K (dashed red), 232 K (green), and 50 K (blue). (b) 5$f_{5/2}$ (blue) and 5$f_{7/2}$ (red) DFT+DMFT projected density of states at 500 K, 232 K, and 50 K.}
\label{fig:fig2_dosdmft}
\end{figure*}

We now turn to the investigation of correlation effects in PuCoGa$_5$ within DFT+DMFT. In Fig.~\ref{fig:fig2_dosdmft} we show the temperature evolution of DFT+DMFT based total, Pu-5$f$, 5$f_{5/2}$, and 5$f_{7/2}$ projected density of states calculated within CTQMC. In comparison with our calculated DFT density of states (see Fig.~\ref{fig:fig1_dosdft}(b)), we find Pu-5$f$ sharp peaks near $E_{f}$ and quite broad peaks at -1 and -1.9 eV, which come mainly from the Co-3d states. These findings are in good agreement with the valence band spectrum of PuCoGa$_5$ obtained from photoemission measurements.~\cite{pes}

Looking at the Pu-5$f$ density of states (lower panel of Fig.~\ref{fig:fig2_dosdmft}(a)), we notice the appearance of a quasiparticle peak (Kondo resonance), mostly of 5$f_{5/2}$ character (see Fig.\ref{fig:fig2_dosdmft}(b)) just below the Fermi level. At 500 K, we start to see the formation of these quasiparticle states, which are enhanced at low temperatures. These findings are a clear signature of the formation of heavy quasiparticles since at low temperatures the Pu-5f electrons strongly hybridize with the surrounding conduction electrons. It is worth mentioning that this feature was observed in early DMFT calculations using the OCA approximation,~\cite{zhu} where the quasiparticle peak was found too sharp due to the overestimation of renormalizations.
In table~I we present the corresponding orbital occupations. Note that for all temperatures n$_{5/2}$ $\approx 4$ and n$_{7/2}$ $\approx 1$.

\begin{table}[!htb]
\label{rutiles_res}
\caption{DFT+DMFT occupancies of 5$f_{5/2}$ and 5$f_{7/2}$ states obtained within CTQMC and OCA impurity solvers.}
\begin{ruledtabular}
\begin{tabular}{ccc}
T(K)   & n$_{5/2}$ & n$_{7/2}$                        \\ \hline 
 &   CTQMC &  \\ \hline 
500    &  4.08  &  1.03         \\   
232    &  4.06  &  1.02          \\ 
50     &  4.02  &  0.98           \\ \hline
 &   OCA  & \\ \hline
500    &  4.00  &  0.99         \\   
232    &  4.01  &  1.00          \\ 
50     &  4.03  &  0.99           \\ 
25     &  4.13  &  1.02 \\
\end{tabular}
\end{ruledtabular}
\end{table}

Furthermore, dynamical correlations lead to the emergence of Hubbard bands at high energies. The upper Hubbard band, which come mainly from 5$f_{7/2}$ states, start to appear at 500 K around 1.1 eV and upshifts to 1.3 eV at 50 K. The lower Hubbard band, mainly due to 5$f_{5/2}$ states, is clearly seen at 232 K and 50 K. At 232 K it is centered at -0.8 eV and upshifts to -0.6 eV at 50 K. Surprisingly, the 5$f_{7/2}$ states, with occupancy close to unit, become gapped at 50 K as can be seen in the lower panel of Fig.~\ref{fig:fig2_dosdmft}(b). 

\begin{figure}[!htb]
\includegraphics[scale=0.5318]{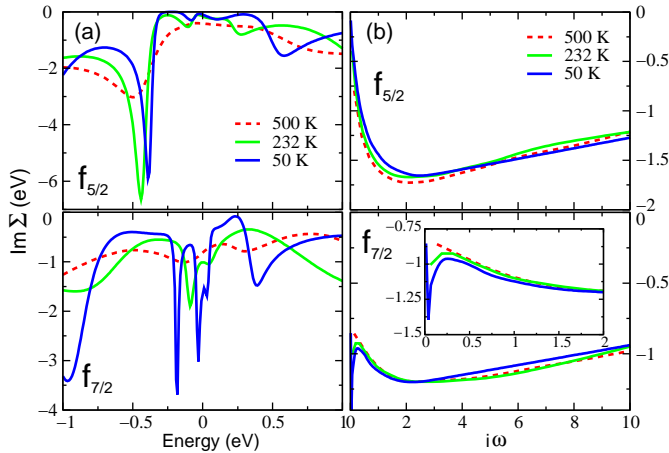}
 \caption{Imaginary part of 5$f_{5/2}$ and 5$f_{7/2}$ self-energies at 500 K (dashed red), 232 K (green), and 50 K (blue) on the (a) real frequency axis , and (b) on the imaginary frequency axis. In the inset we zoom the imaginary part of 5$f_{7/2}$ self-energy in the low energy region.}
\label{fig:self_imag_T}
\end{figure}

Next, we investigate how the dynamical electronic correlations modify the electronic states of PuCoGa$_5$. In Fig.~\ref{fig:self_imag_T}(a)-(b), we show the imaginary part of 5$f_{5/2}$ and 5$f_{7/2}$ components of self-energy for all temperatures considered. 

For temperatures of 500 and 232 K, we observe that the 5$f_{5/2}$ self-energy exhibits a Fermi-liquid like behavior, with a prominent peak at around -0.45 eV as seen in the upper panel of Fig.~\ref{fig:self_imag_T}(a). This high energy feature is also captured in the 5$f_{5/2}$ self-energy computed using the OCA approximation.~\cite{zhu}
As can be seen in Fig.~\ref{fig:self_imag_T}(b), the slope of the imaginary parts at these two temperatures, which is associated with the quasiparticle mass enhancement, is very similar. For the 5$f_{5/2}$ states we estimate a mass enhancement of $\frac{m^*}{m} \approx 5.8$ at 232 K.
At 50 K, the correlations induce a change of behavior in the self-energies. For the 5$f_{5/2}$ states we still observe a Fermi-liquid like behavior, with mass enhancement of $\frac{m^*}{m} \approx 6.4$. However, the imaginary part of 5$f_{7/2}$ states presents two poles at -0.04 and 0.03 eV which are reminiscent of a Mott instability, as can be seen in the lower panel of Fig.~\ref{fig:self_imag_T}(a). We emphasize that the 5$f_{7/2}$ occupancy close to unit, as shown in table~I, favors the appearance of a Mott state.
We mention that the pole below the Fermi level start to appear at 232 K, although in this case it is centered around -0.09 eV.
Looking at this component on the imaginary frequency axis, we find that 5$f_{7/2}$ self-energy exhibits a larger slope than that of 5$f_{5/2}$ component. Moreover, this large slope gives rise to a mass enhancement of $\frac{m^*}{m} \approx 21$, which indicates that the electrons in 5$f_{7/2}$ states are at the edge of a Mott transition. As a result, the 5$f_{7/2}$ projected density of states presents a gap as seen in the lower panel of Fig.~\ref{fig:fig2_dosdmft}(b). Therefore, our DFT+DMFT(CTQMC) calculations suggest the existence of orbital-dependent correlations in PuCoGa$_5$ with substantially differentiation at low temperatures.

\begin{figure}[!htb]
\includegraphics[scale=0.542]{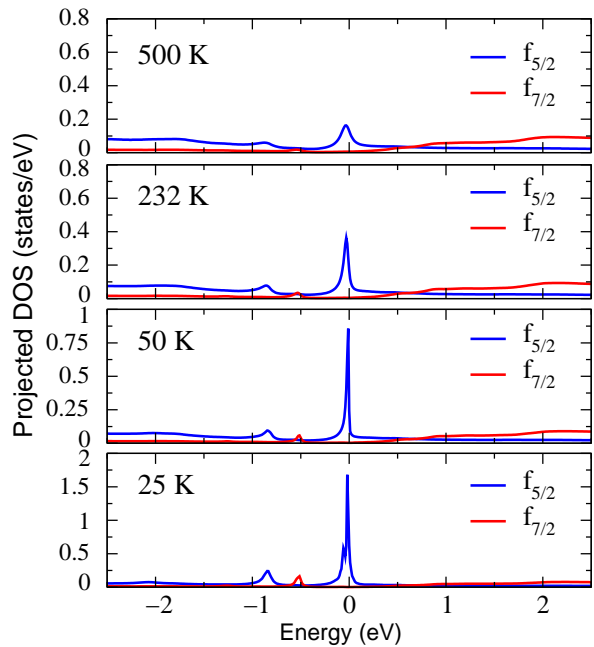}
 \caption{DFT+DMFT(OCA) based Pu-5$f_{5/2}$ (blue) and Pu-5$f_{7/2}$ (red) projected density of states at 500 K, 232 K, 50 K, and 25 K.}
\label{fig:pdos_pu115_oca}
\end{figure}

Another hallmark of orbital-dependent correlations in multiorbital systems is the difference of coherence scales of the orbitals. In order to explore the buildup of coherence in PuCoGa$_5$ we employ the computationally less expensive OCA impurity solver to temperatures down to 25 K. Similar temperature independent orbital occupancies are calculated within this solver as presented in table~I. In Fig.~\ref{fig:pdos_pu115_oca} we display the calculated temperature evolution of the Pu-5$f$ projected density of states from 500 to 25 K. 
As the temperature is reduced we observe the appearance of a quasiparticle peak near the Fermi energy, where the quasiparticle peak height increases upon decrease in temperature. This behavior was also observed in early calculations for the heavy fermion Ce-115 materials within OCA~\cite{shimScience} and is also in agreement with our CTQMC calculations. There are also additional peaks below the Fermi energy which are reminiscent of atomic multiplets observed in the spectra of the $\delta$ phase of elemental Pu.~\cite{shimOCAPu} More important, we find that even at 500 K a quasiparticle peak of 5$f_{5/2}$ character starts to develop whereas there is no sign of Kondo resonance associated with the 5$f_{7/2}$ states for temperatures down to 25 K. 
Furthermore, the 5$f_{7/2}$ spectral function is essentially temperature independent for temperatures down to 25 K, the lowest temperature we could explore before the OCA solver breaks down. Hence the coherence scale of the 5$f_{7/2}$ states is less than 25 K.
This indicates a drastic difference of  Kondo temperatures (T$_K$) of electrons in the 5$f_{5/2}$ and 5$f_{7/2}$ states, where T$_{K}$ of the latter is very small. 
Therefore, our DFT+DMFT(OCA) calculations emphasize the existence of orbital-dependence of correlations in 
PuCoGa$_5$, where the 5$f_{5/2}$ coherence sets in at high temperatures without no sign of Kondo peak for the 5$f_{7/2}$ states down to 25 K.

\section{Conclusions}

In summary, we have performed first-principles calculations at the level of fully charge self-consistent DFT+DMFT to investigate the orbital dependence of correlations in PuCoGa$_5$. From our calculations employing the CTQMC as the impurity solver we find that Pu-5$f$ electrons behave as heavy quasiparticles at low temperatures with strong orbital dependent renormalizations.
Our calculations at 50 K highlight the strongly orbital-dependent correlations in PuCoGa$_5$, wherein electrons in the 5$f_{7/2}$ states are strongly renormalized and are at the edge of a Mott-transition. In addition, our calculations within the OCA demonstrate the orbital differentiation of the coherence energy scales in PuCoGa$_5$, which is hallmark of orbital dependent correlations. Most importantly, our study points towards the universality of the phenomena of orbital differentiation in multiorbital materials. 
It has been conjectured~\cite{mediciPRL} that there is a connection between superconductivity and orbital differentiation. Our discovery of strong orbital differentiation in PuCoGa$_5$, which has the highest T$_c$ of the 5$f$ series, adds an important high temperature superconductor in support of that conjecture. Further microscopic studies are needed to investigate the interplay of the orbital differentiation found here and the spin fluctuations which are present in the family of Pu-based compounds.~\cite{tanmoyPRL}

\section{Acknowledgments}
W.B., S.C., and Y.X. Y. acknowledge support from the Center for Computational Design of Functional Strongly Correlated Materials and Theoretical Spectroscopy.
G.K. was supported by U.S. DOE BES under Grant No. DE-FG02-99ER45761. Many useful discussions with K. Haule are gratefully acknowledged.
An award of computer time was provided by the INCITE program. This research used resources of the Oak Ridge Leadership Computing Facility, which is a DOE Office
of Science User Facility supported under Contract DE-AC05-00OR22725.

\end{document}